\documentclass[11pt]{article}
\usepackage{graphicx}

\def\beq{\begin{eqnarray}}
\def\eeq{\end{eqnarray}}

\def\D{{\cal D}}
\def\S{{\cal S}}
\def\R{\rule{0pt}{15pt}}
\def\T{$\times$}
\def\X{$\!\!\!\!\times$}

\begin{document}

\fontsize{11}{14.5pt}\selectfont

\vspace*{0.4in}

\begin{center}

\begin{center} \Large \bf 
 Representing Numeric Data in 32 Bits \\[4pt] While Preserving 64-Bit Precision
\end{center}

\vspace{8pt}

{\large Radford M. Neal \\[4pt]
  \normalsize Dept.\ of Statistical Sciences and Dept.\ of Computer Science \\
  University of Toronto \\[4pt]
  \texttt{http://www.cs.utoronto.ca/$\sim$radford/}\\
  \texttt{radford@stat.utoronto.ca}\\[4pt]
  8 April 2015}
 
\end{center}

\vspace{20pt}

\noindent Data files often consist of numbers having only a few
significant decimal digits, whose information content would allow
storage in only 32 bits.  However, we may require that arithmetic
operations involving these numbers be done with 64-bit floating-point
precision, which precludes simply representing the data as 32-bit
floating-point values.  Decimal floating point gives a compact and
exact representation, but requires conversion with a slow division
operation before the data can be used in an arithmetic operation.
Here, I show that interesting subsets of 64-bit floating-point values
can be compactly and exactly represented by the 32 bits consisting of
the sign, exponent, and high-order part of the mantissa, with the
lower-order 32 bits of the mantissa filled in by a table lookup
indexed by bits from the part of the mantissa that is retained, and
possibly some bits from the exponent.  For example, decimal data with
four or fewer digits to the left of the decimal point and two or fewer
digits to the right of the decimal point can be represented in this
way, using a decoding table with 32 entries, indexed by the
lower-order 5 bits of the retained part of the mantissa. Data
consisting of six decimal digits with the decimal point in any of the
seven positions before or after one of the digits can also be
represented this way, and decoded using a table indexed by 19 bits
from the mantissa and exponent.  Encoding with such a scheme is a
simple copy of half the 64-bit value, followed if necessary by
verification that the value can be represented, by checking that it
decodes correctly.  Decoding requires only extraction of index bits
and a table lookup.  Lookup in a small table will usually reference
fast cache memory, and even with larger tables, decoding is still
faster than conversion from decimal floating point with a
division operation.  I present several variations on these schemes,
show how they perform on various recent computer systems, and discuss
how such schemes might be used to automatically compress large arrays
in interpretive languages such as R.

\newpage

\subsection*{Introduction}\vspace*{-8pt}

\noindent Numbers that originate from reading text in a data file (as
well as other sources) often have a small number of significant
decimal digits, scaled by a small power of ten.  We may therefore hope
to reduce memory usage, and possibly processing time, by representing
large arrays of such numbers using a 32-bit representation, rather than
the standard 64-bit binary floating point representation (IEEE
Computer Society, 2008) that is used for numbers in programming
languages such as R, and for ``double precision'' numbers in languages
such as C.  However, it is highly desirable that arithmetic involving
such compactly-represented numbers produce exactly the same results as
if they had a 64-bit floating point representation.  This is
particularly important if conversion to a compact representation is
done automatically, as a programming language implementation decision
that is meant to have no visible effect (except, of course, on space
and time usage).

Requiring that numerical results be identical to those using a 64-bit
floating point representation unfortunately eliminates the most
obvious, and very likely fastest, compact representation --- the 32-bit
``single precision'' floating-point representation that is also
standard (IEEE Computer Society, 2008).  Every number with up to seven
significant decimal digits maps to a distinct 32-bit single precision
value, with no information loss.  However, when these single precision
values are converted to 64-bit double precision in the standard
(hardware-supported) way and then used in arithmetic operations, the
results are in general not the same as if a 64-bit floating-point
representation had been used.  The problem is that the standard
conversion by extending the mantissa of a single precision number with
zeros does not produce the correct double precision representation of
a number, such as 0.1, whose binary expansion is non-terminating.

One compact representation that can produce the same results as 64-bit
binary floating point is decimal floating point.  Cowlishaw (2003) has
advanced several reasons for using a decimal floating point format,
such as its ability to distinguish 1.2 from 1.200, but here I consider
it only as a way of exactly representing some 64-bit binary floating
point values in 32 bits.  For this purpose, we might use a 28-bit
signed integer, $M$, and a 4-bit unsigned exponent, $e$, to represent
$M \times 10^{-e}$.  The value $e=15$ might be used to represent a
``NaN'' or ``NA'' value, as needed in R to represent missing values.
Conversion to 64-bit binary floating point can be done by converting
$M$ to binary floating point (exactly), obtaining by table lookup the
(exact) binary floating-point representation of $10^e$ (or a ``NaN''
value when $e=15$), and finally performing a 64-bit binary
floating-point division.  If the decimal floating point value was read
from a data file with eight or fewer significant digits, and its
magnitude is representable within the exponent range of this scheme,
the result of converting it to 64-bit binary floating point should
match the result of directly converting the text in the data file,
since the latter conversion should be done with exactly the same
operations.

Unfortunately, the floating point division operation required to
convert from a decimal floating point representation is quite slow
(compared to other operations) on most current processors.
Multiplication is typically much faster, but in a binary
representation, multiplying by 0.1 is not the same as dividing by 10,
because 0.1 does not have an exact binary floating point
representation. Hardware support for decimal floating point might
speed up this conversion to 64-bit binary floating point somewhat, but
the need for division may limit the improvement possible.  Such
hardware support is in any case not common at present.

In this paper I present a framework for compactly representing subsets
of 64-bit binary floating point values in 32 bits, from which
conversion is faster than from decimal floating point on most current
processors.  The compact 32-bit representation is simply half the
64-bit floating point representation, consisting of the sign,
exponent, and high-order part of the mantissa.  Conversion to the full
64-bit floating-point value is done by filling in the low-order 32
bits of the mantissa from a lookup table, indexed by low-order bits of
the retained part of the mantissa and in some cases by certain bits
from the exponent.  Different versions of this scheme allow for larger
or smaller sets of useful values, with one tradeoff being the size of
the lookup table needed, which affects conversion speed when the table
is too large for the processor's memory cache.  I investigate possible
sets of represented values and associated performance tradeoffs with
several recent processor architectures and compilers, in systems with
varying cache memory sizes and varying processor/memory speed ratios.
Finally, I discuss how such methods might be used in an implementation
of an interpretive language such as R, in order to automatically (and
invisibly) compress large numeric arrays.

C programs for the methods and performance tests are
provided as supplementary information.

\subsection*{Compact representation by 
             half of a 64-bit floating-point value}\vspace*{-8pt}

Although the schemes developed here are motivated by a desire to
compactly represent 64-bit floating point values that are derived from
decimal representations, these schemes can be seen more generally as
compactly representing numbers in a subset, $\S$, of 64-bit
floating-point values that includes some desired subset, which I will
denote by $\D$.

For all these schemes, the compact 32-bit representation of a 64-bit
floating-point value is simply the 32 bits consisting of the sign,
exponent, and high-order mantissa bits of the 64-bit floating-point
value.  I will consider only 64-bit floating-point values in the IEEE
standard format (IEEE Computer Society, 2008), which consist of 1 sign
bit, 11 exponent bits, and 52 mantissa bits, which give 53 effective
bits of precision, since the highest-order bit is alway an implicit 1
(except for denormalized numbers).  The 32-bit compact representation
therefore has 1 sign bit, 11 exponent bits, and 20 mantissa bits
(giving 21 bits of precision).

In all the schemes I consider here, the compact 32-bit value is
converted (decoded) to a 64-bit floating-point value using $m$ bits
from the mantissa part of the compact value, and $e$ (possibly~0) bits
from exponent part, offset from the bottom of the exponent by $f$.
These $m+e$ bits are used to index a table of $2^{m+e}$ possible
values for the lower 32 bits of the mantissa.  The decoded 64-bit
floating-point value consists of the value looked up in this table
together with the 32-bit compact representation.  This procedure is
illustrated here:\vspace{8pt}

\centerline{~\includegraphics[scale=0.47]{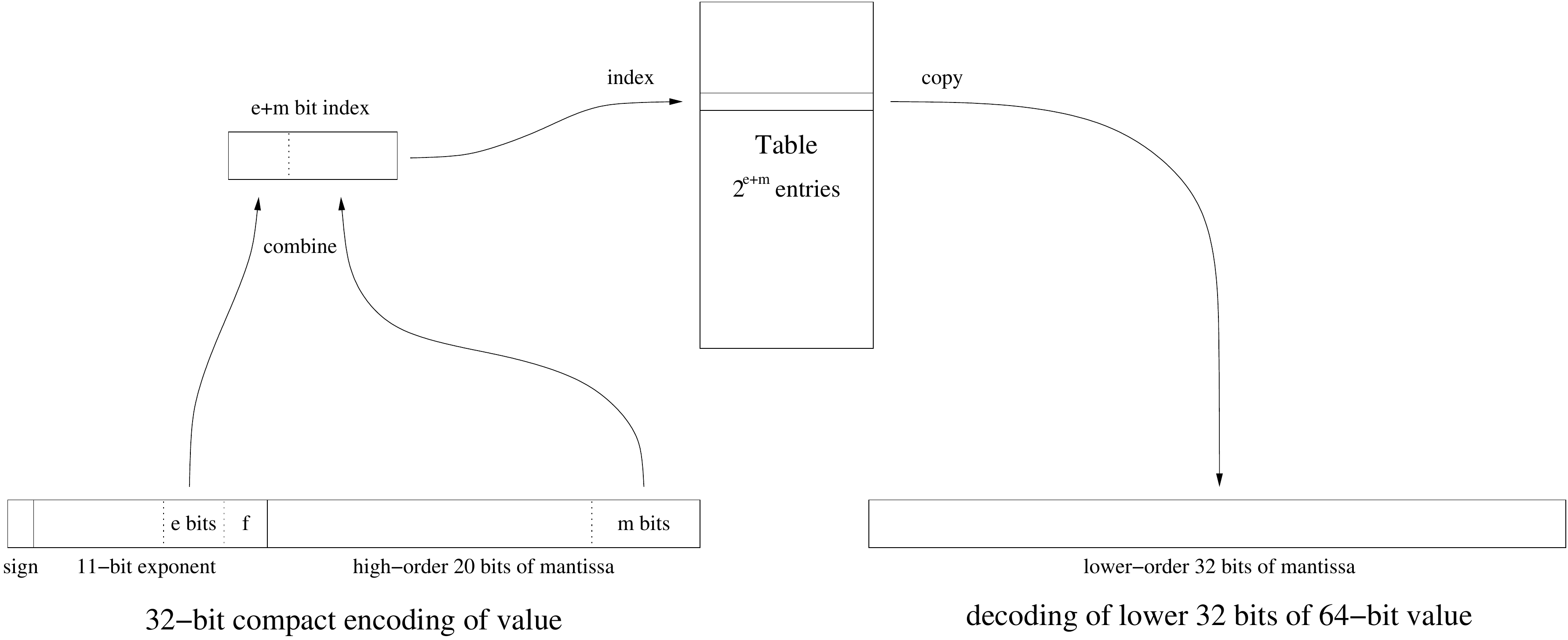}}

\pagebreak

Pseudo-code is shown below for the decoding procedure, which takes as inputs
a 32-bit compact value, $t$, and an array Table with $2^{m+e}$ elements,
indexed starting at 0:
\begin{tabbing}
~~~~~ \= Set $i$ to the integer whose binary representation 
         consists of bits $0\,\ldots\,(m-1)$ of the \\
      \> ~~~~mantissa part of $t$ and bits $f\,\ldots\,(f+e-1)$ of the
             exponent part of $t$. \\[6pt]
      \> Fetch Table[$i$], and concatenate it with $t$ to produce
         the decoded 64-bit floating-point value.
\end{tabbing}
Here, bits are numbered starting with 0 for the lowest-order bit.  
The bits from $t$ combined to produce $i$ may be combined in any consistent 
order, though using the bits from the mantissa as the low-order
bits of the index (as in the illustration above) has implementation advantages.
Note that in some schemes $e$ is zero, in which case the $m$ bits from the 
mantissa are used directly as the index.

Converting a 64-bit value to compact form is done by simply copying
the 32-bit half that constitutes the compact representation.  If the
64-bit value being encoded is not known to be in the set $\S$, the
compact representation is then decoded, the decoded value is compared
to the original 64-bit value, and failure is reported if these values
do not match exactly.

To design a particular scheme in this framework, we need to decide
which bits of the 32-bit compact value are used to index the lookup table
--- that is, choose $m$, $e$, and $f$ --- and what values this lookup
table will contain.  We aim to create a scheme that can represent any
value in some set~$\D$ --- for example, $\D$ might consist of the 2000000
values of the form $\pm$\verb|ddd.ddd|, where each \verb|d| is a decimal digit
from 0 to 9.  The choice of $m$, $e$, and $f$ can be done by trial
and error, perhaps looking for the smallest value of $m+e$ for which a scheme
that can represent all values in $\D$ is possible.  For given $m$, $e$,
and $f$, the procedure below can be used to fill in the lookup
table, or to report failure if representing all values in $\D$ is not possible
(with the chosen values of $m$, $e$, and~$f$):
\begin{tabbing}
~~~~~ \= For $i$ from 0 to $2^{m+e}-1$, set Table[$i$] to 0 (or some other 
         default value). \\[6pt]
      \> For $i$ from 0 to $2^{m+e}-1$, set Used[$i$] to False. \\[6pt]
      \> For each $v$ in $\D$ (in any order): \\[6pt]
      \> ~~~~~ \= Set $t$ to the 32 bits of $v$ containing the sign, the 
                  exponent, and the high-order 20 bits \\
      \>       \> ~~~~of the mantissa. \\[6pt]
      \>       \> Set $i$ to the integer whose binary representation 
                  consists of bits $0\,\ldots\,(m-1)$ of the \\
      \>       \> ~~~~mantissa part of $t$ and bits $f\,\ldots\,(f+e-1)$
                  of the exponent part of $t$. \\[6pt]
      \>       \> If Used[$i$] is True, report failure and stop. \\[6pt]
      \>       \> Otherwise, set Table[$i$] to the 32 lowest-order
                  mantissa bits of $v$, and set Used[$i$] to True.
\end{tabbing}
The way in which bits from $t$ are combined to give $i$ above must match 
the way that will later be used in the decoding procedure.  The output
of this procedure is the array Table, which is then used for decoding.  
The Used array is discarded once
the design procedure has finished.

If this design procedure does not fail, all 64-bit values in $\D$ can
be compactly represented in the scheme designed, and exactly decoded.
In addition, unless all elements of Used are True, some 64-bit
floating-point values that are not in $\D$ will also have compact
representations.  The full set of such values is the set $\S$, of
which $\D$ will be a subset.  Since any 32-bit compact value 
decodes to something, $\S$ will always contain $2^{32}$ members.  The
identity of those members of $\S$ not in $\D$ will depend on the
choice of default value for table entries used in the first step of the
procedure above.

A variation on the decoding procedure that sometimes reduces the space
used for the decoding table is possible when the number of distinct
entries in Table is no more than $2^{16} = 65536$.  We can then use
indirect indexing, replacing the $2^{m+e}$ 32-bit entries in Table
with a table of $2^{m+e}$ 16-bit entries.  The entry obtained from
this table is then used to index a table containing the distinct
32-bit values in the original table.  If the number of distinct
entries in Table is much less than $2^{m+e}$, this will reduce the
memory required by almost a factor of two, and may reduce the time for
decoding if table accesses are then more likely to reference cache
memory.

\subsection*{Some decodable subsets}\vspace*{-8pt}

The utility of this approach to compact representation of 64-bit
floating point values depends crucially on what subsets, $\D$, can be
compactly represented, and for those subsets that can be represented,
on the size of the decoding table, which will have $2^{m+e}$ entries, and
hence occupy $4\times2^{m+e}$ 8-bit bytes (though this may be reduced 
if indirect indexing is used).

Some sets, $\D$, even if they have no more than $2^{32}$ members,
cannot be compactly represented in a scheme of this type, for any
values of $m$, $e$, and $f$.  This will be the case if $\D$ contains
two distinct 64-bit floating-point values for which the 32-bit
portions used as the compact representation are identical.  Other
subsets, $\D$, may be decodable only with a large value for $m+e$.

The NA value used to represent missing data in R poses a problem in
this regard.  In current R implementations, it is a NaN (``Not a
Number'') value with mantissa bits set to the binary representation of
the integer 1954, for which the upper 20 bits are all zero.  With this
choice for the NA value, the 20 bits of the mantissa that are included
in its compact representation are identical to those for the number
zero, so it would not be possible for both NA and zero to be in the
set $\D$ in a scheme with $e=0$.  This issue will also restrict
schemes in which $e>0$.  Fortunately, there appears to be no strong
reason why NA in R could not be changed to a NaN value in which the
upper 20 bits of the mantissa are all ones, and the lower 32 bits are
the binary representation of 1954, and this is the choice for NA that
I will assume below.

Table~\ref{tbl-subsets1} shows several subsets, $\D$, that can be
represented by a scheme in this framework using only mantissa bits for
decoding, along with the required value of $m$ and the resulting table
size.  For these schemes, indirect tables are pointless, since their
total table size is greater than for a direct table.

\begin{table}[b]

\texttt{\small
\begin{center}
\begin{tabular}{llrrrrr}
 &\rm Forms of & $m$ &\rm Table~ &\rm Distinct &\rm Bytes Using \\
 &\rm Numbers & &\rm Entries &\rm Entries$\,$ &\rm Direct Table 
 \\\hline
A & ddddd.d   & 3  &      8~ &     6~ &     32~~ & \\\hline
B & dddd.dd   & 5  &     32~ &    26~ &    128~~ & \\\hline
C & dddd.     & 7  &    128~ &   126~ &    512~~ & \\
  & ddd.ddd \\\hline
D & ddd.d     & 10 &   1024~ &   626~ &   4096~~ & \\
  & dd.dddd \\\hline
E & dd.dd     & 12 &  4096~ &  3126~ &  16384~~ & \\
  & d.ddddd \\\hline
F & dd.       & 14 & 16384~ & 15626~ &  65536~~ & \\
  & d.ddd \\
  & .dddddd \\\hline
\end{tabular}
\end{center}
}

\caption{Some subsets of 64-bit floating-point numbers that can be
represented compactly with decoding based on mantissa bits only (ie,
with $e=0$).  All subsets include the NA value, as well as the numbers
matching the decimal representations shown (plus their negations).
}\label{tbl-subsets1}

\end{table}

\begin{table}[p]

\texttt{\small
\begin{center}
\begin{tabular}{llrrrrrrr}
 &\rm Forms of & $m$ & $e$ & $f$ &\rm Table~ &\rm Distinct &\rm Bytes Using$\,$
 &\rm Bytes Using~~ \\
 &\rm Numbers & & & &\rm Entries &\rm Entries$\,$ &\rm Direct Table 
 &\rm Indirect Tables \\\hline
W & ddddd0.& 10 & 4  & 1 & 16384~ &  626~ & 65536~~ &  35272~~ \\
  & ddddd.d \\
  & dddd.dd \\
  & ddd.ddd \\
  & dd.dddd \\\hline
X & dd0000000.& 10 & 5  & 1 & 32768~ &  9435~ & 131072~~ &  69172~~ \\
  & dd000000. \\
  & dddd000. \\
  & dddddd. \\
  & ddddd.d \\
  & dddd.dd \\
  & ddd.ddd \\
  & dd.dddd \\
  & .000dd \\
  & .0000dd \\
  & .00000dd \\
  & .000000dd \\
  & .0000000dd \\
  & .00000000dd \\
  & .000000000dd \\\hline
Y & d0000000. & 12 & 5  & 1 &131072~ &  5926~ & 524288~~ & 285848~~ \\
  & dddd000. \\
  & dddddd. \\
  & ddddd.d \\
  & dddd.dd \\
  & ddd.ddd \\
  & dd.dddd \\
  & d.ddddd \\
  & .000ddd \\
  & .0000ddd \\
  & .00000ddd \\
  & .000000ddd \\
  & .0000000ddd \\
  & .00000000ddd \\
  & .000000000ddd \\\hline
Z & dd0000000.  & 14 & 5  & 1 &524288~ & 15626~ &2097152~~ &1111080~~ \\
  & ddd00000. \\
  & dddd000. \\
  & dddddd. \\
  & ddddd.d \\
  & dddd.dd \\
  & ddd.ddd \\
  & dd.dddd \\
  & d.ddddd \\
  & .dddddd \\
  & .0000ddd \\
  & .00000ddd \\
  & .000000ddd \\
  & .0000000ddd \\
  & .00000000ddd \\
  & .000000000ddd \\\hline
\end{tabular}
\end{center}
}

\caption{Some subsets of 64-bit floating-point numbers that can be
represented compactly with decoding based on both mantissa bits and
exponent bits.  All subsets include the NA value, as well as
the numbers matching the decimal representations shown (plus their
negations).  
}\label{tbl-subsets2}

\end{table}

Scheme $A$, which can compactly represent any positive or negative
number of the form ddddd.d, such as $12345.6$ or $-888$, requires only
a a very small table of 8 entries.  Schemes $B$ through $F$ represent
subsets that allow for increasing numbers of digits to the right of
the decimal point, but with the tradeoff of fewer digits to the left
of the decimal point, and increased table size.

As shown in Table~\ref{tbl-subsets2}, substantially bigger subsets can
be represented in schemes where $e$ is non-zero, though at the cost of
additional bit manipulation to combine index bits, and larger tables.
For scheme W, the set $\D$ can be described as consisting of any
six-digit number with from one to four of the six digits to the right
of the decimal place, plus integers with six digits of which the last
is zero.  Schemes X, Y, and Z expand the set numbers that can be
represented, including numbers spread over a considerable range of
magnitudes, though with few significant figures, plus numbers having
six digits with the decimal place in various positions (for scheme
Z, before or after any of these six digits).

Note that these schemes can be extended to include additional
useful numbers in their sets $\D$.  For example, scheme Y can be
extended to also include numbers of the form 1dddddd. and 1ddd.ddd,
without any increase in direct or indirect table sizes.  Of course,
this results in some other numbers being removed from the set $\S$,
but those other numbers may not be as useful.

Note also that schemes can be designed for sets $\D$ that are not
defined in terms of decimal representations.  For example, the
64-bit floating-point approximations to all rational numbers of the
form $n/m$ where $n$ is an integer from $-13332$ to $+13332$ and $m$
is an integer from $1$ to $100$ can be represented in a scheme with
$m=13$ and $e=0$.

\subsection*{Software implementation}\vspace*{-8pt}

Some issues arise when implementing encoding and decoding operations
for these schemes in software, particularly when writing in a
high-level language and aiming at portability to different machine
architectures.  Here, I will consider only implementations written in
C, and more specifically, following the revised C99 standard (International
Standards Organization, 2007), in conjunction with the IEEE standard
for floating point arithmetic (IEEE Computer Society, 2008).

The first issue is how to obtain the two 32-bit halves of a 64-bit
floating-point value (a C \verb|double|), which requires bypassing the C
type system.  The C ``union'' construct is designed to allow this, but
its effect may depend on the particular machine architecture being
used.  

One option is to use a union of a \verb|double| field with an array of
two 32-bit unsigned integers, declared as \verb|uint32_t|.  However,
which of the two 32-bit integers contains the upper bits of the 64-bit
double value will depend on whether the machine architecture is
``little-endian'' (for example, the x86 architecture) or
``big-endian'' (for example, the SPARC architecture).  An alternative
that works on both architectures is to use a union of a \verb|double|
field with an 64-bit unsigned integer field declared as
\verb|uint64_t|, and to then access the upper and lower 32-bit halves
of the \verb|double| field by shift and mask operations on the
\verb|uint64_t| field.  In my experience with modern processors and
compilers, the latter approach is also generally faster, and never
consistently slower, so I will adopt it here.  Note that neither
approach is guaranteed to work by the C99 standard, since the
representations of double and unsigned integer values might not be
consistently either big-endian or little-endian, but this does not
appear to be a problem in practice.

When bits from the exponent are used for table lookup when decoding
(that is, $e$ is non-zero), they may be combined with the index bits
from the mantissa in any consistent way.  However, letting the bits
from the mantissa be the low-order bits of the index saves a shift
operation, and is adopted here.  This order is also likely to produce
better cache performance (assuming cache lines are larger than 32
bits), since when a series of values are decoded, their mantissa 
bits are likely to be more variable than their exponent bits.

Finally, it is crucial for good performance that encoding and decoding
operations be performed inline, avoiding function call overhead.
Using modern C compilers, with standard optimizations enabled, this can
be accomplished by defining the small encoding and decoding functions,
declared as \verb|static|~\verb|inline|, within a header file included 
in modules that use them.

\subsection*{Performance on recent computer systems}\vspace*{-8pt}

The speed with which these compact representations can be decoded and
used in arithmetic operations will depend on numerous aspects of the
computer system used, such as the relative speeds of the processor and
memory, the sizes of memory caches, and the degree to which the
processor is capable of overlapping integer and floating-point
operations with each other and with memory accesses.  Here, I will
report results of tests on a variety of fairly recent systems,
providing some illustration of how such factors affect performance.

These tests are all of sequential operations on large vectors (that
is, one-dimensional arrays).  Operations on individual values,
accessed non-sequentially, might show different behaviour.  However,
in the context of an interpretive language implementation, such
operations may in any case be dominated by interpretive overheads.
Also, all the tests performed here use a single execution thread.  It
would be interesting to look at performance when several threads are
executing in parallel, perhaps with more than one accessing the same
compressed data, but a meaningful performance evaluation in such a
context is beyond the scope of this paper.

The compact representation schemes tested here are Scheme C from
Table~\ref{tbl-subsets1} and Schemes X and Z from
Table~\ref{tbl-subsets2}.  For Schemes X and Z, both direct table
lookup and lookup with indirect indexing were tested.  For comparison,
the vector operations were also performed on uncompressed data, and
on data represented in decimal floating point (using 4 bits to specify
a non-negative power of ten and 28 bits for a signed integer scaled by
dividing by that power).

Five operations were tested, each using two data distributions.  As
inputs, these operations take one, two, or three vectors of data, in
one of the compact representations (or represented as uncompressed
64-bit floating point values for comparison). They produce as output a
vector of uncompressed values, or a single uncompressed value, as
follows:

\begin{tabular}{ll}
\textbf{copy}       & copy of the input \\[3pt]
\textbf{vector sum} & sum of all values in the input vector \\[3pt] 
\textbf{scalar $\times$ vector} 
                    & product of 123.456789 times the input vector \\[3pt] 
\textbf{vector $+$ vector} & vector sum of two input vectors \\[3pt] 
\textbf{linear combination} & linear combination of three input vectors
                      with coefficients 1.1, 2.2, and 3.3
\end{tabular}

\noindent These operations were programmed in C, with inlined
functions used for decoding the compressed representations.  A high
optimization level was used when compiling, with options to enable
code generation optimized for the particular processor on which the
code was run.  The code, scripts, and output for the tests are
available as supplemental information.

For the tests reported, all input vectors contained 3,000,000 randomly
generated values.  Two distributions for these values were tried.  In
the first, each value has the form \verb|ddd.ddd|, with each \verb|d|
being a random digit from~0 to~9.  In the second, the forms
\verb|dd.dddd|, \verb|ddd.ddd|, and \verb|dddd.dd| were cycled
through, with each \verb|d| again being random.  Only the first
distribution was used for Scheme C, since it cannot represent all
values generated with the second distribution.  Each operation, on
data generated from each distribution, was repeated 100 times, and the
total time was recorded in seconds (to three decimal places, though
the last digit is often not meaningful due to random time variation).


\begin{table}[p]

\begin{center}\begin{tabular}{|l|rrrrr|rrrrr|}\hline
 & \multicolumn{5}{c|}{Data:\ \ ddd.ddd} 
 & \multicolumn{5}{c|}{Data:\ \ dd.dddd, ddd.ddd, dddd.dd} \\[-1pt]
 & \multicolumn{1}{c}{\small} 
 & \multicolumn{1}{c}{\small vector}
 & \multicolumn{1}{c}{\small $\!\!\!$scalar $\times\!\!\!\!\!$}
 & \multicolumn{1}{c}{\small $\!$vector $+\!\!$}
 & \multicolumn{1}{c}{\small linear}
 & \multicolumn{1}{|c}{\small} 
 & \multicolumn{1}{c}{\small vector}
 & \multicolumn{1}{c}{\small $\!\!\!$scalar $\times\!\!\!\!\!$}
 & \multicolumn{1}{c}{\small $\!$vector $+\!\!$}
 & \multicolumn{1}{c|}{\small linear} \\[-4pt]
 & \multicolumn{1}{c}{\small copy}
 & \multicolumn{1}{c}{\small sum}
 & \multicolumn{1}{c}{\small vector}
 & \multicolumn{1}{c}{\small $\!\!$vector}
 & \multicolumn{1}{c}{\small comb.}
 & \multicolumn{1}{|c}{\small copy}
 & \multicolumn{1}{c}{\small sum}
 & \multicolumn{1}{c}{\small vector}
 & \multicolumn{1}{c}{\small $\!\!$vector}
 & \multicolumn{1}{c|}{\small comb.} \\\hline
\small $\!$Uncompressed\rule{0pt}{15pt}
                             & 0.490 & 0.310 & 0.470 & 0.650 & 0.860
                             & 0.500 & 0.300 & 0.470 & 0.660 & 0.870 \\\hline
\small $\!$Scheme C:\R       & \T0.86& \T1.26& \T1.02& \T1.11& \T1.42 &&&&& \\
\small $\!$~~~direct table   & 0.420 & 0.390 & 0.480 & 0.720 & 1.220
                             & --~~~ & --~~~ & --~~~ & --~~~ & --~~~ \\\hline
\small $\!$Scheme X:\R       & \T1.12& \T1.81& \T1.40& \T1.60& \T1.92
                             & \T1.12& \T1.87& \T1.45& \T1.62& \T1.94\\
\small $\!$~~~direct table   & 0.550 & 0.560 & 0.660 & 1.040 & 1.650
                             & 0.560 & 0.560 & 0.680 & 1.070 & 1.690 \\
\small \R                    & \T1.16& \T1.94& \T1.51& \T1.77& \T2.13
                             & \T1.16& \T2.00& \T1.51& \T1.76& \T2.10 \\
\small $\!$~~~indirect table & 0.570 & 0.600 & 0.710 & 1.150 & 1.830
                             & 0.580 & 0.600 & 0.710 & 1.160 & 1.830 \\\hline
\small $\!$Scheme Z:\R       & \T1.63& \T2.42& \T2.02& \T2.26& \T2.69
                             & \T1.90& \T2.93& \T2.32& \T2.58& \T3.03\\
\small $\!$~~~direct table   & 0.800 & 0.750 & 0.950 & 1.470 & 2.310
                             & 0.950 & 0.880 & 1.090 & 1.700 & 2.640 \\
\small \R                    & \T1.55& \T2.29& \T1.91& \T2.20& \T2.57
                             & \T2.02& \T3.07& \T2.45& \T2.82& \T3.29\\
\small $\!$~~~indirect table & 0.760 & 0.710 & 0.900 & 1.430 & 2.210
                             & 1.010 & 0.920 & 1.150 & 1.860 & 2.860 \\\hline
\small \R                    & \T4.61& \T7.55& \T5.77& \T6.12& \T7.98
                             & \T4.50& \T7.80& \T5.77& \T6.03& \T7.89\\
\small $\!$Decimal float     & 2.260 & 2.340 & 2.710 & 3.980 & 6.860
                             & 2.250 & 2.340 & 2.710 & 3.980 & 6.860 \\\hline
\end{tabular}\end{center}

\vspace{5pt}

\begin{center}\small \begin{tabular}{rl}
Processor: & Intel X5680, 6 cores, 3.33 GHz, launched 2010 \\
RAM:       & 24 GBytes, DDR3 1333 MHz \\
Caches:    & 32 KByte I \& D L1, 256 KByte L2, 12 MByte L3 (shared) \\
Software:  & Ubuntu 12.04.5, gcc 4.8.1
\end{tabular}\end{center}\vspace{-3pt}

\caption{Performance on a Dell Precision T7500 high-end workstation.
}\label{tbl-results-dell}

\end{table}

\begin{table}[p]

\begin{center}\begin{tabular}{|l|rrrrr|rrrrr|}\hline
 & \multicolumn{5}{c|}{Data:\ \ ddd.ddd} 
 & \multicolumn{5}{c|}{Data:\ \ dd.dddd, ddd.ddd, dddd.dd} \\[-1pt]
 & \multicolumn{1}{c}{\small} 
 & \multicolumn{1}{c}{\small vector}
 & \multicolumn{1}{c}{\small $\!\!\!$scalar $\times\!\!\!\!\!$}
 & \multicolumn{1}{c}{\small $\!$vector $+\!\!$}
 & \multicolumn{1}{c}{\small linear}
 & \multicolumn{1}{|c}{\small} 
 & \multicolumn{1}{c}{\small vector}
 & \multicolumn{1}{c}{\small $\!\!\!$scalar $\times\!\!\!\!\!$}
 & \multicolumn{1}{c}{\small $\!$vector $+\!\!$}
 & \multicolumn{1}{c|}{\small linear} \\[-4pt]
 & \multicolumn{1}{c}{\small copy}
 & \multicolumn{1}{c}{\small sum}
 & \multicolumn{1}{c}{\small vector}
 & \multicolumn{1}{c}{\small $\!\!$vector}
 & \multicolumn{1}{c}{\small comb.}
 & \multicolumn{1}{|c}{\small copy}
 & \multicolumn{1}{c}{\small sum}
 & \multicolumn{1}{c}{\small vector}
 & \multicolumn{1}{c}{\small $\!\!$vector}
 & \multicolumn{1}{c|}{\small comb.} \\\hline
\small $\!$Uncompressed\rule{0pt}{15pt}
                             & 1.879 & 0.710 & 1.718 & 2.302 & 2.639 
                             & 1.859 & 0.713 & 1.697 & 2.282 & 2.617 \\\hline
\small $\!$Scheme C:\R       & \T0.63& \T2.14& \T0.87& \T1.10& \T1.66&&&&& \\
\small $\!$~~~direct table   & 1.180 & 1.518 & 1.490 & 2.541 & 4.359
                             & --~~~ & --~~~ & --~~~ & --~~~ & --~~~ \\\hline
\small $\!$Scheme X:\R       & \T0.91& \T2.80& \T1.19& \T1.72& \T2.23
                             & \T1.03& \T2.99& \T1.32& \T1.92& \T2.47\\
\small $\!$~~~direct table   & 1.705 & 1.988 & 2.040 & 3.949 & 5.858
                             & 1.923 & 2.130 & 2.243 & 4.374 & 6.467 \\
\small \R                    & \T0.96& \T3.19& \T1.27& \T1.82& \T2.42
                             & \T1.01& \T3.21& \T1.32& \T1.89& \T2.50\\
\small $\!$~~~indirect table & 1.812 & 2.267 & 2.179 & 4.196 & 6.372
                             & 1.870 & 2.289 & 2.239 & 4.321 & 6.545 \\\hline
\small $\!$Scheme Z:\R       & \T1.36& \T4.18& \T1.62& \T2.49& \T3.16
                             & \T1.75& \T4.56& \T2.07& \T2.96& \T3.66\\
\small $\!$~~~direct table   & 2.557 & 2.971 & 2.787 & 5.740 & 8.299 
                             & 3.262 & 3.251 & 3.515 & 6.756 & 9.588 \\
\small \R                    & \T1.61& \T4.78& \T2.02& \T2.96& \T3.77
                             & \T2.12& \T5.83& \T2.56& \T3.74& \T4.70\\
\small $\!$~~~indirect table & 3.028 & 3.392 & 3.468 & 6.814 & 9.904
                             & 3.938 & 4.160 & 4.345 & 8.538 &12.309 \\\hline
\small \R                    & \T3.86&\X12.75& \T6.43& \T6.27&\X10.11
                             & \T3.89&\X12.70& \T6.50& \T6.34&\X10.10\\
\small $\!$Decimal float
                             & 7.255 & 9.055 &11.046 &14.438 &26.579
                             & 7.237 & 9.058 &11.037 &14.476 &26.427 \\\hline
\end{tabular}\end{center}

\vspace{5pt}

\begin{center}\small \begin{tabular}{rl}
Processor: & AMD E1-2500, 2 cores, 1.4 GHz, launched 2013 \\
RAM:       & 4 GBytes, DDR3L 1333 MHz \\
Caches:    & 32 KByte I \& D L1, 512 KByte L2 \\
Software:  & Ubuntu 14.04.1, gcc 4.8.2
\end{tabular}\end{center}\vspace{-3pt}

\caption{Performance on a Gateway SX2185 low-end desktop computer.
}\label{tbl-results-gateway}

\end{table}


\begin{table}[p]

\begin{center}\begin{tabular}{|l|rrrrr|rrrrr|}\hline
 & \multicolumn{5}{c|}{Data:\ \ ddd.ddd} 
 & \multicolumn{5}{c|}{Data:\ \ dd.dddd, ddd.ddd, dddd.dd} \\[-1pt]
 & \multicolumn{1}{c}{\small} 
 & \multicolumn{1}{c}{\small vector}
 & \multicolumn{1}{c}{\small $\!\!\!$scalar $\times\!\!\!\!\!$}
 & \multicolumn{1}{c}{\small $\!$vector $+\!\!$}
 & \multicolumn{1}{c}{\small linear}
 & \multicolumn{1}{|c}{\small} 
 & \multicolumn{1}{c}{\small vector}
 & \multicolumn{1}{c}{\small $\!\!\!$scalar $\times\!\!\!\!\!$}
 & \multicolumn{1}{c}{\small $\!$vector $+\!\!$}
 & \multicolumn{1}{c|}{\small linear} \\[-4pt]
 & \multicolumn{1}{c}{\small copy}
 & \multicolumn{1}{c}{\small sum}
 & \multicolumn{1}{c}{\small vector}
 & \multicolumn{1}{c}{\small $\!\!$vector}
 & \multicolumn{1}{c}{\small comb.}
 & \multicolumn{1}{|c}{\small copy}
 & \multicolumn{1}{c}{\small sum}
 & \multicolumn{1}{c}{\small vector}
 & \multicolumn{1}{c}{\small $\!\!$vector}
 & \multicolumn{1}{c|}{\small comb.} \\\hline
\small $\!$Uncompressed\rule{0pt}{15pt}
                             & 1.617 & 0.479 & 1.611 & 2.030 & 2.643
                             & 1.618 & 0.503 & 1.608 & 2.034 & 2.651 \\\hline
\small $\!$Scheme C:\R       & \T0.88& \T1.26& \T0.87& \T0.80& \T0.79&&&&& \\
\small $\!$~~~direct table   & 1.419 & 0.604 & 1.403 & 1.633 & 2.080
                             & --~~~ & --~~~ & --~~~ & --~~~ & --~~~ \\\hline
\small $\!$Scheme X:\R       & \T0.87& \T1.76& \T0.89& \T0.88& \T1.02
                             & \T0.87& \T1.71& \T0.88& \T0.89& \T1.03\\
\small $\!$~~~direct table   & 1.412 & 0.845 & 1.429 & 1.794 & 2.701 
                             & 1.409 & 0.860 & 1.418 & 1.802 & 2.731 \\
\small \R                    & \T0.88& \T1.86& \T0.89& \T0.93& \T1.11
                             & \T0.87& \T1.79& \T0.90& \T0.93& \T1.11\\
\small $\!$~~~indirect table & 1.415 & 0.893 & 1.437 & 1.884 & 2.943
                             & 1.408 & 0.901 & 1.443 & 1.889 & 2.939 \\\hline
\small $\!$Scheme Z:\R       & \T0.94& \T2.06& \T0.95& \T1.11& \T1.31
                             & \T1.01& \T2.04& \T1.02& \T1.24& \T1.36\\
\small $\!$~~~direct table   & 1.520 & 0.988 & 1.528 & 2.262 & 3.462
                             & 1.639 & 1.027 & 1.647 & 2.514 & 3.602 \\
\small \R                    & \T0.93& \T2.32& \T0.97& \T1.19& \T1.52
                             & \T1.05& \T2.48& \T1.10& \T1.42& \T1.71\\
\small $\!$~~~indirect table & 1.503 & 1.095 & 1.555 & 2.421 & 4.030
                             & 1.703 & 1.246 & 1.767 & 2.883 & 4.532 \\\hline
\small \R                    & \T2.45& \T8.44& \T2.45& \T3.87& \T4.53
                             & \T2.43& \T8.03& \T2.47& \T3.88& \T4.52\\
\small $\!$Decimal float
                             & 3.960 & 4.042 & 3.947 & 7.863 &11.980
                             & 3.931 & 4.041 & 3.979 & 7.901 &11.970 \\\hline
\end{tabular}\end{center}

\vspace{5pt}

\begin{center}\small \begin{tabular}{rl}
Processor: & Intel Core 2 Duo (T7700), 2 cores, 2.4 GHz, launched 2007 \\
RAM:       & 4 GBytes, DDR2 667 MHz \\
Caches:    & 32 KByte I \& D L1, 4 MBytes L2 (shared?) \\
Software:  & Mac OS X 10.10.2 (Yosemite), clang 600.0.56
\end{tabular}\end{center}\vspace{-3pt}

\caption{Performance on an Apple MacBook Pro 3,1 laptop computer.
}\label{tbl-results-macbookpro}

\end{table}

\begin{table}[p]

\begin{center}\begin{tabular}{|l|rrrrr|rrrrr|}\hline
 & \multicolumn{5}{c|}{Data:\ \ ddd.ddd} 
 & \multicolumn{5}{c|}{Data:\ \ dd.dddd, ddd.ddd, dddd.dd} \\[-1pt]
 & \multicolumn{1}{c}{\small} 
 & \multicolumn{1}{c}{\small vector}
 & \multicolumn{1}{c}{\small $\!\!\!$scalar $\times\!\!\!\!\!$}
 & \multicolumn{1}{c}{\small $\!$vector $+\!\!$}
 & \multicolumn{1}{c}{\small linear}
 & \multicolumn{1}{|c}{\small} 
 & \multicolumn{1}{c}{\small vector}
 & \multicolumn{1}{c}{\small $\!\!\!$scalar $\times\!\!\!\!\!$}
 & \multicolumn{1}{c}{\small $\!$vector $+\!\!$}
 & \multicolumn{1}{c|}{\small linear} \\[-4pt]
 & \multicolumn{1}{c}{\small copy}
 & \multicolumn{1}{c}{\small sum}
 & \multicolumn{1}{c}{\small vector}
 & \multicolumn{1}{c}{\small $\!\!$vector}
 & \multicolumn{1}{c}{\small comb.}
 & \multicolumn{1}{|c}{\small copy}
 & \multicolumn{1}{c}{\small sum}
 & \multicolumn{1}{c}{\small vector}
 & \multicolumn{1}{c}{\small $\!\!$vector}
 & \multicolumn{1}{c|}{\small comb.} \\\hline
\small $\!$Uncompressed\rule{0pt}{15pt}
                             & 1.580 & 0.543 & 1.568 & 2.041 & 2.636
                             & 1.609 & 0.546 & 1.623 & 2.068 & 2.718 \\\hline
\small $\!$Scheme C:\R       & \T0.87& \T1.29& \T0.88& \T0.80& \T0.89&&&&& \\
\small $\!$~~~direct table   & 1.370 & 0.702 & 1.381 & 1.639 & 2.333
                             & --~~~ & --~~~ & --~~~ & --~~~ & --~~~ \\\hline
\small $\!$Scheme X:         & \T0.88& \T1.87& \T0.91& \T0.97& \T1.21
                             & \T0.87& \T1.89& \T0.89& \T0.98& \T1.20\\
\small $\!$~~~direct table   & 1.391 & 1.015 & 1.429 & 1.976 & 3.183
                             & 1.396 & 1.030 & 1.442 & 2.018 & 3.270 \\
\small \R                    & \T0.89& \T1.98& \T0.93& \T1.04& \T1.32
                             & \T0.88& \T1.98& \T0.90& \T1.03& \T1.29\\
\small $\!$~~~indirect table & 1.402 & 1.077 & 1.456 & 2.120 & 3.492
                             & 1.409 & 1.080 & 1.463 & 2.138 & 3.512 \\\hline
\small $\!$Scheme Z:         & \T0.98& \T2.19& \T1.03& \T1.25& \T1.57
                             & \T1.04& \T2.26& \T1.07& \T1.28& \T1.54\\
\small $\!$~~~direct table   & 1.548 & 1.187 & 1.610 & 2.550 & 4.141
                             & 1.675 & 1.236 & 1.742 & 2.647 & 4.195 \\
\small \R                    & \T0.99& \T2.45& \T1.08& \T1.36& \T1.77
                             & \T1.17& \T2.85& \T1.23& \T1.51& \T1.86\\
\small $\!$~~~indirect table & 1.557 & 1.331 & 1.699 & 2.779 & 4.666
                             & 1.876 & 1.556 & 1.996 & 3.123 & 5.054 \\\hline
\small \R                    & \T1.95& \T5.64& \T1.97& \T3.00& \T3.58
                             & \T1.91& \T5.60& \T1.90& \T2.96& \T3.47\\
\small $\!$Decimal float
                             & 3.084 & 3.060 & 3.084 & 6.125 & 9.428
                             & 3.081 & 3.060 & 3.087 & 6.126 & 9.434 \\\hline
\end{tabular}\end{center}

\vspace{5pt}

\begin{center}\small \begin{tabular}{rl}
Processor: & Intel Core 2 Due (P7350), 2 GHz, launched 2008 \\
RAM:       & 2 GBytes, DDR3 1066 MHz \\
Caches:    & 32 KByte I \& D L1, 3 MBytes L2 (shared?) \\
Software:  & Mac OS X 10.10.2 (Yosemite), clang 600.0.56
\end{tabular}\end{center}\vspace{-3pt}

\caption{Performance on an Apple Mac mini 3,1 mid-range desktop computer.
}\label{tbl-results-macmini}

\end{table}


\begin{table}[p]

\begin{center}\begin{tabular}{|l|rrrrr|rrrrr|}\hline
 & \multicolumn{5}{c|}{Data:\ \ ddd.ddd} 
 & \multicolumn{5}{c|}{Data:\ \ dd.dddd, ddd.ddd, dddd.dd} \\[-1pt]
 & \multicolumn{1}{c}{\small} 
 & \multicolumn{1}{c}{\small vector}
 & \multicolumn{1}{c}{\small $\!\!\!$scalar $\times\!\!\!\!\!$}
 & \multicolumn{1}{c}{\small $\!$vector $+\!\!$}
 & \multicolumn{1}{c}{\small linear}
 & \multicolumn{1}{|c}{\small} 
 & \multicolumn{1}{c}{\small vector}
 & \multicolumn{1}{c}{\small $\!\!\!$scalar $\times\!\!\!\!\!$}
 & \multicolumn{1}{c}{\small $\!$vector $+\!\!$}
 & \multicolumn{1}{c|}{\small linear} \\[-4pt]
 & \multicolumn{1}{c}{\small copy}
 & \multicolumn{1}{c}{\small sum}
 & \multicolumn{1}{c}{\small vector}
 & \multicolumn{1}{c}{\small $\!\!$vector}
 & \multicolumn{1}{c}{\small comb.}
 & \multicolumn{1}{|c}{\small copy}
 & \multicolumn{1}{c}{\small sum}
 & \multicolumn{1}{c}{\small vector}
 & \multicolumn{1}{c}{\small $\!\!$vector}
 & \multicolumn{1}{c|}{\small comb.} \\\hline
\small $\!$Uncompressed\rule{0pt}{15pt}
                             & 4.190 & 4.700 & 4.960 & 8.950 &13.650
                             & 4.180 & 4.700 & 4.960 & 8.950 &13.660 \\\hline
\small $\!$Scheme C:\R       & \T1.38& \T1.25& \T2.52& \T1.75& \T1.76&&&&& \\
\small $\!$~~~direct table   & 5.790 & 5.860 &12.490 &15.660 &24.080
                             & --~~~ & --~~~ & --~~~ & --~~~ & --~~~ \\\hline
\small $\!$Scheme X:\R       & \T2.19& \T1.89& \T2.62& \T2.52& \T2.49
                             & \T2.56& \T2.15& \T2.94& \T2.88& \T2.11\\
\small $\!$~~~direct table   & 9.170 & 8.880 &13.020 &22.560 &34.010
                             &10.710 &10.090 &14.570 &25.780 &38.840 \\
\small \R                    & \T5.06& \T4.41& \T2.74& \T2.55& \T2.50
                             & \T5.66& \T4.92& \T3.27& \T3.14& \T3.08\\
\small $\!$~~~indirect table &21.210 &20.710 &13.570 &22.830 &34.170
                             &23.670 &23.130 &16.210 &28.060 &42.110 \\\hline
\small $\!$Scheme Z:\R       & \T2.88& \T2.37& \T3.22& \T3.20& \T3.16
                             & \T2.94& \T2.40& \T3.26& \T3.24& \T3.20\\
\small $\!$~~~direct table   &12.080 &11.150 &15.950 &28.680 &43.160
                             &12.280 &11.260 &16.150 &29.000 &43.650 \\ 
\small \R                    & \T6.01& \T5.25& \T3.61& \T3.50& \T3.46
                             & \T6.38& \T5.56& \T3.93& \T3.86& \T3.80\\
\small $\!$~~~indirect table &25.190 &24.670 &17.910 &31.350 &47.170
                             &26.680 &26.110 &19.470 &34.560 &51.950 \\\hline
\small \R                    & \T4.18& \T3.92& \T3.78& \T3.83& \T3.90
                             & \T4.19& \T3.91& \T3.78& \T3.83& \T3.90\\
\small $\!$Decimal float
                             &17.500 &18.420 &18.750 &34.250 &53.240 
                             &17.510 &18.400 &18.750 &34.240 &53.240 \\\hline
\end{tabular}\end{center}

\vspace{5pt}

\begin{center}\small \begin{tabular}{rl}
Processors:& Two UltraSPARC T2 Plus processors, 8 cores each,
             8 threads per core, 1.2 GHz, launched 2008 \\
RAM:       & 16 GBytes, DDR2 667 MHz \\
Caches:    & 16 KBytes I L1, 8 KBytes D L1, 4 MBytes L2 (shared) \\
Software:  & Solaris 11.2, Oracle Solaris Studio 12.4
\end{tabular}\end{center}\vspace{-3pt}

\caption{Performance on a Sun T5140 high-end multithread-optimized server.
}\label{tbl-results-T5140}

\end{table}

\begin{table}[p]

\begin{center}\begin{tabular}{|l|rrrrr|rrrrr|}\hline
 & \multicolumn{5}{c|}{Data:\ \ ddd.ddd} 
 & \multicolumn{5}{c|}{Data:\ \ dd.dddd, ddd.ddd, dddd.dd} \\[-1pt]
 & \multicolumn{1}{c}{\small} 
 & \multicolumn{1}{c}{\small vector}
 & \multicolumn{1}{c}{\small $\!\!\!$scalar $\times\!\!\!\!\!$}
 & \multicolumn{1}{c}{\small $\!$vector $+\!\!$}
 & \multicolumn{1}{c}{\small linear}
 & \multicolumn{1}{|c}{\small} 
 & \multicolumn{1}{c}{\small vector}
 & \multicolumn{1}{c}{\small $\!\!\!$scalar $\times\!\!\!\!\!$}
 & \multicolumn{1}{c}{\small $\!$vector $+\!\!$}
 & \multicolumn{1}{c|}{\small linear} \\[-4pt]
 & \multicolumn{1}{c}{\small copy}
 & \multicolumn{1}{c}{\small sum}
 & \multicolumn{1}{c}{\small vector}
 & \multicolumn{1}{c}{\small $\!\!$vector}
 & \multicolumn{1}{c}{\small comb.}
 & \multicolumn{1}{|c}{\small copy}
 & \multicolumn{1}{c}{\small sum}
 & \multicolumn{1}{c}{\small vector}
 & \multicolumn{1}{c}{\small $\!\!$vector}
 & \multicolumn{1}{c|}{\small comb.} \\\hline
\small $\!$Uncompressed\rule{0pt}{15pt}
                             & 5.066 & 1.969 & 5.069 & 7.670 &10.012
                             & 5.057 & 1.970 & 5.058 & 7.671 & 9.917 \\\hline
\small $\!$Scheme C:\R       & \T1.30& \T1.21& \T1.31& \T0.96& \T1.12&&&&& \\
\small $\!$~~~direct table   & 6.578 & 2.384 & 6.624 & 7.396 &11.224
                             & --~~~ & --~~~ & --~~~ & --~~~ & --~~~ \\\hline
\small $\!$Scheme X:\R       & \T1.35& \T1.87& \T1.46& \T1.42& \T1.53
                             & \T1.40& \T2.06& \T1.54& \T1.54& \T1.67\\
\small $\!$~~~direct table   & 6.831 & 3.673 & 7.405 &10.864 &15.293
                             & 7.104 & 4.068 & 7.812 &11.836 &16.537 \\
\small \R                    & \T1.39& \T2.22& \T1.41& \T1.46& \T2.11
                             & \T1.45& \T2.44& \T1.49& \T1.63& \T2.31\\
\small $\!$~~~indirect table & 7.025 & 4.362 & 7.162 &11.171 &21.077
                             & 7.333 & 4.803 & 7.537 &12.528 &22.929 \\\hline
\small $\!$Scheme Z:\R       & \T1.89& \T3.73& \T2.19& \T2.29& \T2.49
                             & \T2.41& \T4.56& \T2.83& \T2.89& \T3.16\\
\small $\!$~~~direct table   & 9.600 & 7.346 &11.089 &17.550 &24.934
                             &12.164 & 8.992 &14.301 &22.159 &31.314 \\
\small \R                    & \T2.13& \T5.50& \T2.24& \T2.90& \T4.03
                             & \T2.65& \T6.98& \T2.77& \T3.56& \T5.06\\
\small $\!$~~~indirect table &10.805 &10.839 &11.371 &22.242 &40.300
                             &13.422 &13.758 &13.995 &27.315 &50.187 \\\hline
\small \R                    & \T1.46& \T3.43& \T1.67& \T1.88& \T2.15
                             & \T1.46& \T3.43& \T1.67& \T1.88& \T2.17\\
\small $\!$Decimal float
                             & 7.397 & 6.758 & 8.452 &14.400 &21.537
                             & 7.400 & 6.758 & 8.456 &14.401 &21.538 \\\hline
\end{tabular}\end{center}

\vspace{5pt}

\begin{center}\small \begin{tabular}{rl}
Processor: & ARM v7 Cortex-A9, 4 cores, 1 GHz, launched 2008 \\
RAM:       & 2 GBytes, DDR3 1066 MHz \\
Caches:    & 32 KByte I \& D L1, 1 MBytes L2 (shared) \\
Software:  & Debian Linux, gcc 4.8.2
\end{tabular}\end{center}\vspace{-3pt}

\caption{Performance on a Cubox i.MX6 miniature computer.
}\label{tbl-results-cubox}

\end{table}


The results for six computer systems are reported in
Tables~\ref{tbl-results-dell} to~\ref{tbl-results-cubox}. These tables
give the times in seconds.  For methods other than use of uncompressed
data, the times have above them the ratio (preceded by``$\times$'') of
the time below to the corresponding time using uncompressed data.  A
ratio less than one indicates that for this combination of operation,
data distribution, and compression scheme, operating on compressed
data is faster than using uncompressed data.  A ratio greater than one
indicates the reverse, that operating on the compressed data is
slower.

Tables \ref{tbl-results-dell} and \ref{tbl-results-gateway} show
performance on a high-end workstation and a low-end desktop computer,
both of fairly recent vintage, both using x86 architecture processors
(in 64-bit mode).  Although the low-end system is roughly three times
slower, the relative performances of the various representations are
broadly similar for these two systems.  This is perhaps surprising,
given that the high-end system has 12 MBytes of cache, versus 512
KBytes for the low-end system, which one might think would make a
substantial difference at least for the schemes with large decoding
tables.  However, the processor for the low-end system is
substantially slower, while the memory is of the same type, so the
low-end system has faster memory in relation to processor speed, which
perhaps compensates for its smaller cache.

For the systems in Tables \ref{tbl-results-dell} and
\ref{tbl-results-gateway}, Scheme C performs about the same as using
uncompressed data --- slightly faster for some operations, somewhat
slower for others.  This scheme therefore seems quite attractive when
its set of representable values is sufficient.  Scheme X provides a
much larger set of useful values, but (using a direct table) it is
somewhat slower than Scheme C, though it is still mostly less than a
factor of two slower than using uncompressed data.  Using indirect
tables with Schemes X is slightly slower than using a direct table,
though it is possible that the smaller amount of memory required for
indirect tables will provide a compensating advantage in the context
of an actual application, in which other operations are also done.
Scheme Z provides a still larger set of useful values, but at the cost
of a further moderate slowdown.

Schemes X and Z are slightly slower when the data is drawn from the
second distribution rather than from the first distribution,
presumably because the greater variety of data values causes accesses
to the decoding tables to be more spread out, making memory caching
less effective.

Since, on both of these systems, Schemes X and Z are slower than using
uncompressed data, a decision to use one of them must be based on the
advantage of decreased memory usage, the significance of which will
depend on the wider context of the application and the system on which
it runs.  However, if we suppose that reducing memory usage is
necessary, we can see that all of schemes C, X, and Z give faster
performance than using decimal floating point, by a factor of more
than three for Scheme X.  One of these schemes should therefore be
preferred when the set of values that they can represent is
sufficient.

Tables \ref{tbl-results-macbookpro} and \ref{tbl-results-macmini} show
results for two systems using variations of the Intel Core 2 Duo x86
architecture processor (in 64-bit mode), which are of somewhat earlier
vintage than the processors used in the systems of
Tables~\ref{tbl-results-dell} and~\ref{tbl-results-gateway}.  The
MacBook Pro system has a 2.4 GHz processor and 667 MHz memory.  The
Mac mini system has a 2.0 GHz processor (of slightly more recent
design) and 1066 MHz memory.  The results for this pair of systems
therefore illustrate how the ratio of processor speed to memory speed
affects performance.

On both these systems, Scheme C is faster than using uncompressed data
for all operations except the vector sum.  Scheme X using a direct
table is also often faster than using uncompressed data, and Scheme Z
using a direct table is mostly less than a factor of two slower than
using uncompressed data.  The performance (relative to using
uncompressed data) of these compression schemes is more favourable for
both these systems than for the somewhat newer systems of
Tables~\ref{tbl-results-dell} and~\ref{tbl-results-gateway}, though it
is unclear whether the difference is due to any systematic change in
more recent processor designs.  

The performance of Schemes X and Z is somewhat better on the MacBook
Pro than on the Mac mini, in line with expectations of the effect of
the higher processor/memory speed ratio for the MacBook~Pro.  The
magnitude of the difference is perhaps less than would be expected
from naively assessing this speed ratio using clock rates, however.

For both the MacBook Pro and the Mac mini, Schemes C, X, and Z are all
considerably faster than using decimal floating point.  Decimal
floating point is relatively better on the Mac mini than on the
MacBook Pro, presumably due to some improvement in the floating point
division method (or in its ability to overlap with other operations),
but it remains slower than Scheme X (with a direct table) by factors
of between two and three for the various operations

Tables \ref{tbl-results-T5140} and \ref{tbl-results-cubox} show
results on two systems whose processors do not use the (very common)
x86 architecture.  The Sun T5140 uses T2 Plus processors that
implement the SPARC V9 architecture in a manner designed to support
many concurrent threads (128 for the dual-processor T5140), by use of
within-core multithreading.  It is intended for use as a server that
processes a large volume of transactions handled in multiple threads;
its single thread performance is slower than earlier SPARC systems
that did not have within-core multithreading.  The Cortex-A9
implements the ARM v7 architecture.  It is used in many tablet
computers and smart phones, though the Cubox system tested here is a
miniature computer that can be used as a desktop computer or home
entertainment system.

On the T2 Plus processor in the T514, the interleaving of many threads
within one processor core produces the effect that each thread sees
memory as being relatively faster compared to the speed with which the
thread executes instructions than would the case in a processor
without such interleaving.  One would expect that the advantage of
using a compressed representation that comes from a reduction in the
amount of data transferred would then be less, at least in a test such
as this in which only a single thread is executing.  And indeed, we
see that unlike the x86 systems in Tables~\ref{tbl-results-dell}
to~\ref{tbl-results-macmini}, operations using Scheme~C are always
slower than when using uncompressed data, by at least a factor of
1.25.  The relative performances of Schemes~X and~Z are also worse
than for those systems.  However, these schemes (with direct tables)
are nevertheless faster than decimal floating point (though by fairly
small factors).

The Cubox system's Cortex-A9 processor has a low 1~GHz clock rate, but
reasonably fast 1066 MHz memory, which may explain why the performance
relative to using uncompressed data of Schemes~C, X, and~Z is
generally worse than for the four x86 systems (though better than for
the T5140).  Puzzlingly, though, the relative performance of decimal
floating point is better on this system than on any of the others,
though it is still a bit slower than Scheme~X.

\subsection*{Automatic use in an interpretive programming language
  implementation}\vspace*{-8pt}

My principal motivation for investigating these compact representation
schemes is their possible use in an interpretive implementation of a
language such as R (R Core Team, 2015).  The goal is for the
implementation to switch automatically to a compact representation for
large numeric vectors, matrices, or other objects when such a compact
representation is possible, with this switch having no visible effect,
apart from its impact on speed and memory usage.  In particular, I
hope in future to include some version of such a scheme in the pqR
implementation of R (Neal, 2013--2015).

Use of a compact representation scheme could instead be implemented
as a visible option, which could be explicitly requested by an
application program, after which it would be responsible for
explicitly handling encoding and decoding operations.  One can also
imagine using these compact representations for other purposes, such as
reducing the space needed to store large databases, either
automatically or as an explicit option.  In this paper, however, I
will consider only use of these schemes for automatic compression of
objects stored in main memory.

In this context, the intent is that application programs will not be
aware of whether numbers are represented in some compact scheme.  The
only exceptions to this that I envision are that there might be some
way for an application program to request that an object be
converted to a compact representation, if this is possible and has not
already been done automatically, and there might also be a way of querying
whether an object is compactly represented, purely as an aid for
understanding performance issues.

Using a compact representation is worthwhile only when an object is
large, and is possible then only when there is some compact coding
scheme within which all numerical values in the object are
representable.  Run-time checks for whether an object is compactly
represented will therefore generally be necessary.  In a language
implemented by compilation to native machine code, such checks would
be present in numerous compiled code segments, and the overhead of
these checks in both execution time and code size could be
significant.  Here, I consider only interpretive implementations, in
which these checks will be present only in a limited number of places
in the interpreter, and the time they take may be small compared to
other interpretive overheads.

Nevertheless, modifying every code segment in the interpreter that
accesses numerical values in order to add checks for whether these
numbers are compactly represented would be arduous.  Furthermore, R
defines an interface by which an application program can call
functions written in C, Fortran, or another language, which may access
numeric data that could also be compactly represented.  If existing
code is to continue to work, it is necessary to provide a default
interface in which compactly represented data is never seen.  Selected
parts of the interpreter could then be modified to be aware of compact
representations, and to implement operations using them without
conversion.  The application program interface might also be extended
to provide a way for user-written code to manipulate data that is
compactly represented, while still allowing existing user code to be
used without change, since objects would automatically be converted to
the standard representation if they are passed to such code.

In the pqR implementation of R, a similar issue already arises with
objects containing numerical values that may not have been computed
yet, either because the computation is being done, or might be done,
in some other concurrently-executing thread, or because the
computation has been deferred in the hope that it may later be merged
with a subsequent computation.  This is handled in pqR by making the
default versions of functions that return an object --- such as by
fetching the value of a variable, or by evaluating an expression ---
wait for any computation on the object to finish before returning it.
Code segments that are prepared to handle objects whose computation is
pending can, however, call alternative versions of these functions,
which will not wait for computations to finish.

A similar approach could be used to handle compactly represented
objects.  Code that is not aware of compact representations will call
the default functions, which will automatically convert any compact
object to its standard format.  Code that is prepared to operate on
compactly represented data can call alternative versions of these
functions that do not convert compact data.  With this approach, the
use of compact representations will be beneficial only if they are
mostly manipulated by code that is aware of such representations, so
that they can remain in their compact representation, but correct
operation does not depend on every code segment in the interpreter and
in application-program C code being modified to deal with compact
representations.

Conversion from a compact representation to the standard representation
will also be necessary if a numeric element in the object is replaced
by one that is not in the set, $\S$, of values that can be represented
in the scheme being used.

The obvious way to convert from a compact representation to the standard
representation is to allocate new space to hold the converted object,
decode the compact representation to the standard representation, and
then free the space that held the compact representation.  

This approach has several problems, however.  First, the amount of
memory required will temporarily be 50\% larger than if the standard
representation had been used from the beginning.  Second, it is not
clear what to do if the object is shared --- for example, is the value
of several variables.  In R, such sharing is done only to save space,
and R semantics require that a copy be made when one value is
modified, so the other will remain unchanged.  It is unclear whether
in this context it would be best to replace all references to the
compact representation by the converted representation, or only the
reference for which the standard representation is now required.  If
all references are replaced, they would have to be found, which is not
easy.  But if only one reference is replaced, memory for both
representations will remain allocated.  Finally, decoding objects to
newly-allocated space changes the address of the object, and also
could trigger a garbage collection at the point where an operation
prompting the conversion is done.  Existing code may have been written
in a way that assumes such changes in address and initiations of
garbage collection do not happen at times when they now could.

A different approach therefore seems to be required, and is possible
if the operating system supports allocation of space within the
virtual address space of the process without physical memory being
allocated until the addresses are actually referenced.  Current Linux,
Solaris, and Mac OS X operating systems (and likely others as well)
have such a facility, and use it in the malloc function in the C library
when a large block of memory is requested.

In this approach, when an object is created, space sufficient to hold
its standard representation is always reserved for it in the virtual
address space.  If a compact representation is used initially, only
the first half of this space will be referenced, and hence the second
half will not occupy any physical memory.  If later the object needs
to be converted to the standard representation, it can be decoded in
place (starting at the end), at which time physical memory will be
assigned to the second half.  The address of an object stays the same
when it is converted, and garbage collection will not be triggered by
the conversion operation, since no additional (virtual) memory is
allocated.  This makes handling automatic conversion to standard
format quite analogous to the deferred evaluation mechanism that is
already implemented in pqR.

Similarly, when an object that is currently in the standard
representation is converted to a compact representation, its virtual
address and virtual storage allocation can remain the same, while the
physical memory that is no longer required after conversion can be
released (if the operating system supports this).

This approach to implementing compact representations will produce the
desired benefits as long as physical memory is the limiting resource,
with the available virtual address space being larger than physical
memory. (A virtual address space a factor of two larger than physical
memory should always be sufficient, and in some applications a smaller
factor might suffice.)  The virtual address space can be expected to
be sufficiently large in 64-bit architectures (although considerations
other than pointer size might limit it to less than $2^{64}$ bytes).
Systems with a 32-bit address space and no more than 2 GBytes of
physical memory should also benefit from using compact
representations.  However, in a system that combines a large physical
memory with a limited 32-bit address space, this approach to
implementing compact representations may produce no benefit.

An unusual aspect of the framework for compact representations in this
paper is that although there are many different schemes, such as seen
in Tables~\ref{tbl-subsets1} and~\ref{tbl-subsets2}, the compact
representation of a number is identical for every such scheme.
Because of this, a compactly represented object can be noted as being
encoded by not just one such scheme, but by a set of schemes.
Decoding according to any of these schemes will produce the same
result, so whichever scheme has the fastest decoder can be used.  When
a numerical value in the object is changed, any schemes that cannot
represent this new value will be removed from the set of schemes that
can decode the object.  If this set becomes empty, the object will
have to be converted to the standard, non-compact representation.

Note in particular that when creating an object from data read from a
file, compact representations of numbers from the file can be stored
as they are read, even though the set of schemes that can be used for
decoding these representations will not be known until all data has
been read.  If a number is read that cannot be represented in any of
the schemes that are implemented, the compact representations of
previously read numbers can be expanded, and the standard
representation used for the remaining data.

\subsection*{Acknowledgements}\vspace*{-8pt}

This research was supported by Natural Sciences and Engineering
Research Council of Canada. The author holds a Canada Research Chair
in Statistics and Machine Learning.

\subsection*{References}\vspace*{-8pt}

\leftmargini 0.2in
\labelsep 0in

\begin{description}
\itemsep 2pt

\item
Cowlishaw, M.~F.\ (2003) ``Decimal Floating-Point: Algorism for Computers'',
  in \textit{Proceedings of the 16th IEEE Symposium on Computer Arithmetic}.

\item
  IEEE Computer Society\ (2008) \textit{IEEE Standard for Floating-Point
  Arithmetic}.

\item International Standards Organization (2007) Combined 
document with C99 + TC1 + TC2 + TC3,
available from \verb|http://www.open-std.org/jtc1/sc22/wg14/www/docs/n1256.pdf|

\item Neal, R.~M.\ (2013--2015) pqR --- a pretty quick version of R,
\verb|http://pqR-project.org|

\item R Core Team (2015) \textit{R Language Definition (Draft)}, available
at \verb|http://r-project.org|

\end{description}

\end{document}